\documentclass[journal=jctcce,manuscript=article]{achemso} 
\usepackage{graphicx} 
\usepackage{amsmath} 
\usepackage{mhchem} 
\usepackage{longtable} 
\usepackage{cleveref} 
\usepackage{subfigure} 
\usepackage{subfig} 

\newcommand{\newcite}[1]{\scalebox{1.3}[1.33]{\raisebox{-0.85ex}{\cite{#1}}}} 

\Crefname{equation}{Eq.}{Eqs.} 

\author{Chia-Nan Yeh} 
\affiliation{Department of Physics, National Taiwan University, Taipei 10617, Taiwan} 

\author{Jeng-Da Chai} 
\email{jdchai@phys.ntu.edu.tw} 
\affiliation{Department of Physics, National Taiwan University, Taipei 10617, Taiwan} 
\alsoaffiliation{Center for Theoretical Sciences and Center for Quantum Science and Engineering, National Taiwan University, Taipei 10617, Taiwan}

\title{Role of Kekul\'{e} and Non-Kekul\'{e} Structures in the Radical Character of Alternant Polycyclic Aromatic Hydrocarbons: A TAO-DFT Study} 

\begin{document} 

\setlength{\fboxrule}{0 pt}

\begin{abstract} 

We investigate the role of Kekul\'{e} and non-Kekul\'{e} structures in the radical character of alternant polycyclic aromatic hydrocarbons (PAHs) using thermally-assisted-occupation density functional 
theory (TAO-DFT), an efficient electronic structure method for the study of large ground-state systems with strong static 
correlation effects. Our results reveal that the studies of Kekul\'{e} and non-Kekul\'{e} structures qualitatively describe the radical character of alternant PAHs, which could be useful when electronic 
structure calculations are infeasible due to the expensive computational cost. In addition, our results support previous findings on the increase in radical character with increasing system size. For 
alternant PAHs with the same number of aromatic rings, the geometrical arrangements of aromatic rings are responsible for their radical character. 

\end{abstract}

\section*{Introduction} 

Polycyclic aromatic hydrocarbons (PAHs), which consist of three or more aromatic rings, are molecules made of carbon and hydrogen atoms only. While some PAHs possess non-radical (i.e., 
closed-shell) character, many others can exhibit radical (i.e., open-shell or multi-reference) character due to the delocalized nature of $\pi$-electrons. PAHs with multi-reference ground states, which 
belong to strongly correlated systems, have been actively studied, owing to their attractive properties and applications \cite{OS1,OS2,OS4,OS6,OS7,OS8,OS9,OS10,OS11,OS12,OS13}. 

In particular, alternant PAHs, which are PAHs constituted only of six-membered rings, are planar molecules. Consequently, alternant PAHs have been frequently taken as finite-size models of 
hydrogen-passivated graphene nanoribbons and nanoflakes. Experimentally, in spite of the promising properties of alternant PAHs, their instability is a commonly encountered problem. Because of 
their high reactivity, alternant PAHs with substantial radical character are usually short-lived. Methods for stabilization are necessary for their practical use. Theoretically, due to the multi-reference 
character of ground-state wavefunctions, alternant PAHs with significant radical character cannot be adequately described by conventional electronic structure methods \cite{GNR_new1,Acene-Carter,
HM,FG2,FG3,Acene-DMRG,Acene-Jiang,NOON1,TAO-DFT,GNRs-DMRG,GNRs-PHF,GNRs-MRAQCC,TAO-GGA,GNRs-MRAQCC2,GNRs-TAO,n-PP,PC5,PC6,PC7}, including the very popular 
Kohn-Sham density functional theory (KS-DFT) \cite{KS-DFT2} with commonly used semilocal \cite{PBE}, hybrid \cite{hybrid,LCHirao,wB97X,wB97X-D,wM05-D,LC-D3}, and 
double-hybrid \cite{B2PLYP,wB97X-2,PBE0-2,SCAN0-2} density functionals. 

High-level {\it ab initio} multi-reference methods \cite{Acene-DMRG,GNRs-DMRG,GNRs-MRAQCC,GNRs-MRAQCC2,2-RDM,NOON1,multi-reference} are typically required to accurately describe 
the electronic properties of alternant PAHs with pronounced radical character, and their radical character may be examined from the analysis of the natural orbital occupation numbers (NOONs) [i.e., 
the eigenvalues of one-electron reduced density matrix] \cite{NOON,NOON1,NOON2,Acene-DMRG,GNRs-DMRG,GNRs-MRAQCC,GNRs-MRAQCC2}. For a molecule with a singlet ground state, 
if some of the NOONs deviate strongly from the closed-shell values of two/zero, the molecule exhibits pronounced radical character (e.g., diradical, tetraradical, or higher-order radical character). 
On the other hand, if all the NOONs are close to the closed-shell values of two/zero, the molecule has non-radical character. 

\begin{figure}[H] 
\includegraphics[scale=0.7]{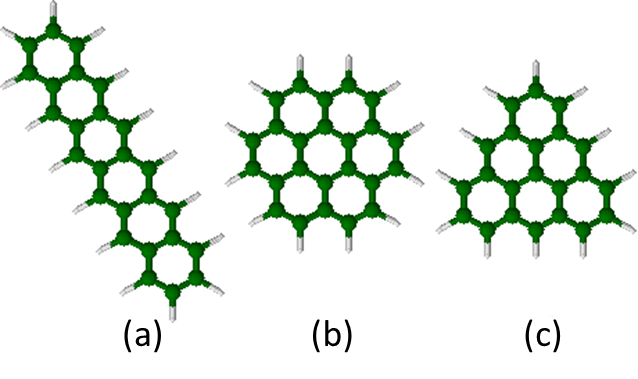} 
\caption{\label{example} 
Geometries of (a) hexacene, (b) superbenzene, and (c) triangulene.} 
\end{figure}

\Cref{example} shows three alternant PAHs: hexacene, superbenzene, and triangulene, all of which consist of six aromatic rings. To assess their radical character, Pelzer \emph{et al.} \cite{NOON1} 
calculated the occupation numbers of the highest occupied ($n_\text{HONO}$) and lowest unoccupied ($n_\text{LUNO}$) natural orbitals for the lowest singlet states of these molecules, using the 
active-space variational two-electron reduced-density-matrix (RDM-CASSCF) method \cite{2-RDM}. Their results revealed that triangulene ($n_\text{HONO}=1.095$ and $n_\text{LUNO}=0.916$) 
exhibits strong radical character, hexacene possesses partial radical character ($n_\text{HONO}=1.542$ and $n_\text{LUNO}=0.472$), and superbenzene ($n_\text{HONO}=1.818$ and 
$n_\text{LUNO}=0.204$) has a stable singlet ground state. As these molecules all contain six aromatic rings, it seems plausible that the geometrical arrangements of aromatic rings may be 
responsible for the radical character of alternant PAHs. Arguments in support of this are also available in several studies \cite{2-RDM,Acene-DMRG,GNRs-DMRG,GNRs-PHF,GNRs-MRAQCC,
GNRs-MRAQCC2,GNRs-TAO,PC5,PC6,PC7}. However, owing to the prohibitively high cost of accurate multi-reference calculations, there have been very few studies on large alternant PAHs (e.g., 
containing up to a few thousand electrons). 

In order to extend the investigation to include large alternant PAHs, an efficient electronic structure method for the study of large ground-state systems with strong static correlation effects should be 
adopted. Recently, we have developed thermally-assisted-occupation density functional theory (TAO-DFT) \cite{TAO-DFT,TAO-GGA}, which can be an ideal electronic structure method for the study 
of large alternant PAHs \cite{GNRs-TAO}. Besides, the orbital occupation numbers from TAO-DFT, which are intended to simulate the NOONs, could provide information useful in assessing the possible 
radical character of alternant PAHs. 

In this work, we first show that the orbital occupation numbers from TAO-DFT are qualitatively similar to the NOONs from RDM-CASSCF on a set of small alternant PAHs. Next, the orbital occupation 
numbers from TAO-DFT are adopted to examine the radical character of large alternant PAHs. In addition, we also propose a simple model based on the extended Clar's rule to qualitatively describe 
the radical character of alternant PAHs (i.e., without performing electronic structure calculations), which is especially desirable for very large alternant PAHs. From this simple model, we can easily 
illustrate how the geometrical arrangements of aromatic rings are responsible for the radical character of the alternant PAHs with the same number of aromatic rings.

\section*{Simple Model Based on the Extended Clar's Rule} 

\begin{figure}[H] 
\includegraphics[scale=0.7]{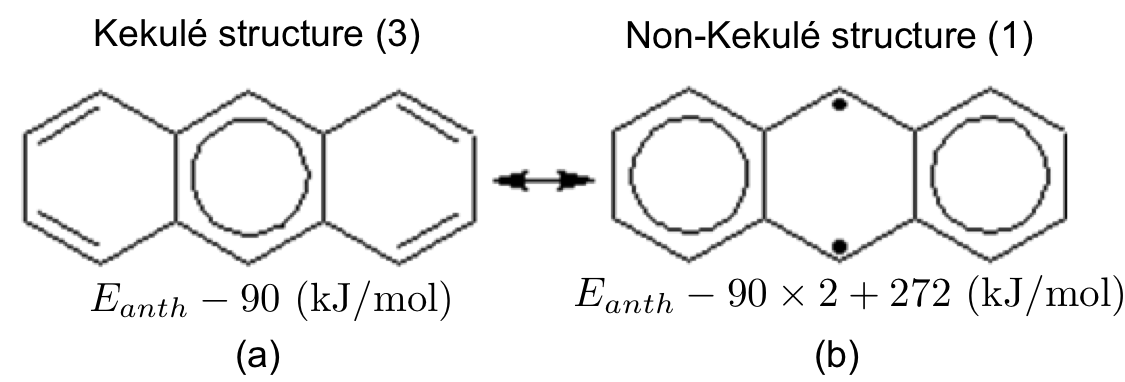} 
\caption{\label{resonance} 
Most important Kekul\'{e} and non-Kekul\'{e} structures of anthracene. 
Here the aromatic sextets are marked with circles, and the unpaired $\pi$-electrons are marked with black dots. 
The energies and degeneracies (in parentheses) of the structures are shown.} 
\end{figure}

The bonding patterns in benzene can be understood by the resonance between the two Kekul\'{e} structures of benzene \cite{Kekule}. The superposition of these two structures, Clar's aromatic 
sextet \cite{Clar1}, can be interpreted as six $\pi$-electrons moving all around the aromatic ring. For alternant PAHs, more than two Kekul\'{e} structures may, however, be needed to describe the 
resonance. Clar's rule states that the Kekul\'{e} structure with the largest number of disjoint aromatic sextets (i.e., benzene-like moieties), the Clar structure, is the most important structure for the 
characterization of the properties of PAHs \cite{Clar1,Clar2,Clar5,Clar6}. However, the studies of Kekul\'{e} structures alone cannot adequately describe open-shell singlet states. 
Therefore, recent studies have extended Clar's rule to include the non-Kekel\'{e} structures (e.g., the structures with unpaired $\pi$-electrons) \cite{NK1,NK2,NK3}. Accordingly, to describe the 
resonance in alternant PAHs, all the Kekul\'{e} and non-Kekul\'{e} structures can be adopted (see \Cref{resonance}). For alternant PAHs with strong radical character, the non-Kekel\'{e} structures are 
expected to be essential. 

Nevertheless, in the extended Clar's rule, the relationship between the non-Kekul\'{e} structures and radical character of alternant PAHs remains unclear. In previous studies, it has been suggested 
that the analysis of the radical character of alternant PAHs could be focused solely on the energy balance between the break of a covalent bond and aromatic stabilization \cite{NK1,NK2,NK3}. In 
other words, only the energies of the Kekul\'{e} and non-Kekul\'{e} structures have been previously considered. In this work, we highlight the importance of the degeneracies of the Kekul\'{e} and 
non-Kekul\'{e} structures in the radical character of alternant PAHs. 

\begin{figure}[H] 
\includegraphics[scale=0.7]{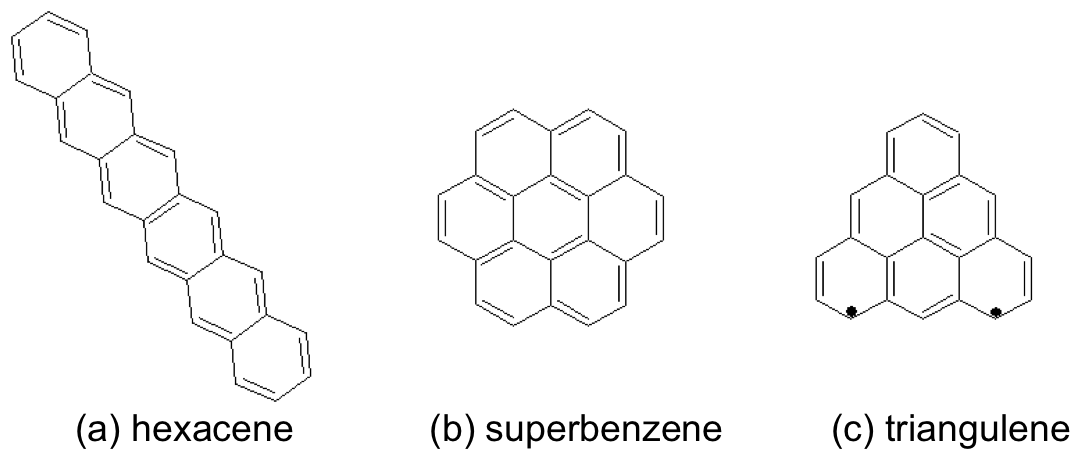} 
\caption{\label{fig:S1} 
Standard structures of (a) hexacene, (b) superbenzene, and (c) triangulene.} 
\end{figure}

Here we propose a simple model to qualitatively describe the radical character of alternant PAHs. 
To start with, we define the standard structure of molecule $\alpha$ as the structure: 
\begin{enumerate} 
\item Which has no aromatic sextet 
\item Whose $\pi$-electrons pair with their nearest neighbors to form carbon-carbon (C-C) double bonds, yielding the least number of unpaired $\pi$-electrons. 
\end{enumerate} 
\Cref{fig:S1} illustrates the standard structures of (a) hexacene, (b) superbenzene, and (c) triangulene. 
Unlike hexacene and superbenzene, there are two unpaired $\pi$-electrons in the standard structure of triangulene, owing to the peculiar geometry of triangulene. 

Next, we assign the energies of the structures of molecule $\alpha$. The energy of the standard structure is assigned as $E_{\alpha}$ (kJ/mol). 
As it has been reported that the aromatic stabilization energy of benzene is about 90 kJ/mol \cite{DE2}, and the C-C $\pi$-bonding energy is about 272 kJ/mol \cite{pi_bonding}, 
the energy of the structure with one extra aromatic sextet is assigned as $E_{\alpha} - 90$ (kJ/mol), 
the energy of the structure with two extra aromatic sextets is assigned as $E_{\alpha} - 90 \times 2$ (kJ/mol), and so on, while 
the energy of the structure with two extra unpaired $\pi$-electrons is assigned as $E_{\alpha} + 272$ (kJ/mol), 
the energy of the structure with four extra unpaired $\pi$-electrons is assigned as $E_{\alpha} + 272 \times 2$ (kJ/mol), and so on. 
By counting the number of extra aromatic sextets and the number of extra unpaired $\pi$-electrons in a structure, the energy of the structure can be assigned. 
For brevity, the structures with the lowest, second lowest, and third lowest energies are referred as the most important, second most important, and third most important structures, respectively. 

\begin{figure}[H] 
\includegraphics[scale=0.6]{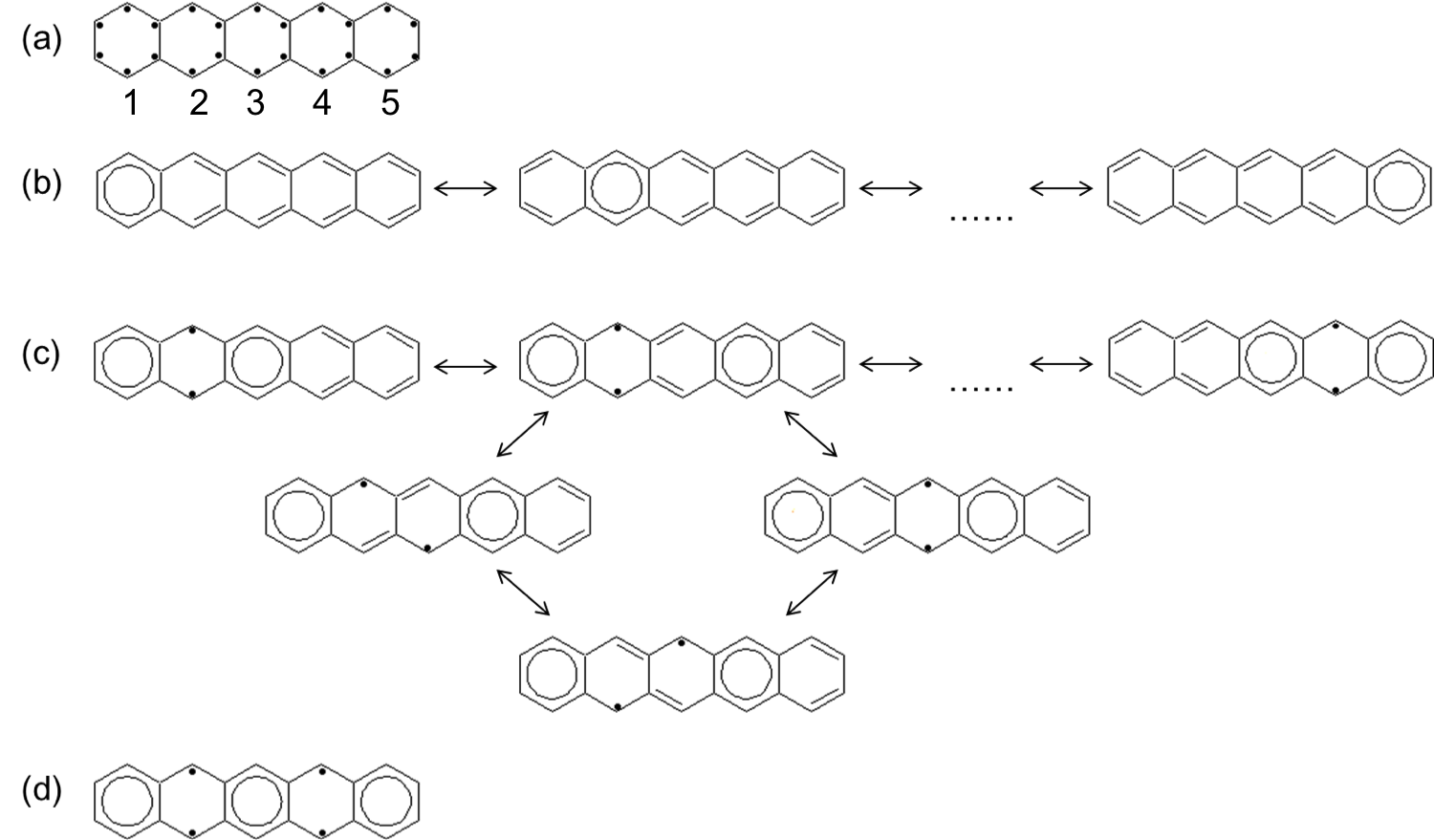} 
\caption{\label{fig:S2} 
Pentacene with 
(a) all the $\pi$-electrons unpaired (marked with black dots), where the aromatic rings are numbered. 
(b) the most important structures 
(c) the second most important structures 
(d) the third most important structure.} 
\end{figure}

In addition, we count the degeneracies of the structures of molecule $\alpha$. 
Note that the structures with the same energy (e.g., those with the same number of extra aromatic sextets and the same number of extra unpaired $\pi$-electrons) are degenerate. 
Here we demonstrate how to count the degeneracies of the top three most important structures of pentacene (see \Cref{fig:S2}) for instance. 
\Cref{fig:S2}(a) shows all the $\pi$-electrons unpaired (marked with black dots), which may later pair with their nearest neighbors to form C-C double bonds or aromatic sextets, or remain unpaired. 
\Cref{fig:S2}(b) shows various ways to draw the most important structures (i.e., the structures with one extra aromatic sextet). Clearly, the aromatic sextet can be placed in one of the five aromatic rings. 
Noteworthy, once the aromatic sextet is placed, there is only one way for the remaining $\pi$-electrons pairing with their nearest neighbors (i.e., forming C-C double bonds). 
Therefore, the degeneracy of the most important structure is 5. 
\Cref{fig:S2}(c) shows the second most important structures (i.e., the structures with two extra aromatic sextets and two extra unpaired $\pi$-electrons). 
Since the two aromatic sextets should be placed in two of the five aromatic rings, and they cannot be adjacent, there are five ways to arrange them. 
Besides, the two unpaired $\pi$-electrons should be placed accordingly. For example, 
if the two aromatic sextets are placed in rings 1 and 4, the two unpaired $\pi$-electrons can only be placed at ring 2 or 3, and they cannot both be placed at the top or bottom. 
As a result, there are only four ways to arrange these two unpaired $\pi$-electrons. 
Therefore, the degeneracy of the second most important structure is 20. 
\Cref{fig:S2}(d) shows the only way to draw the third most important structure (i.e., the structure with three extra aromatic sextets and four extra unpaired $\pi$-electrons). 
As the three aromatic sextets cannot be adjacently placed, they can only be placed in rings 1, 3, and 5. 
Therefore, the degeneracy of the third most important structure is 1. 

Finally, similar to expanding a wavefunction by a linear combination of configuration state functions in the full configuration interaction method, in our simple model, 
the radical character of molecule $\alpha$ is argued to be related to all the possible Kekul\'{e} and non-Kekul\'{e} structures as well as their corresponding weights. 
The weight of a structure should be dependent on the energy and degeneracy of the structure. 
The structure with lower energy and greater degeneracy should have a larger weight, and hence govern the radical character of molecule $\alpha$ more significantly, while 
the structure with higher energy and less degeneracy should have a smaller weight, and hence govern the radical character of molecule $\alpha$ less significantly.

\section*{Computational Details} 

As alternant PAHs are all planar molecules, we define a one-dimensional (1D) alternant PAH as the alternant PAH with aromatic rings connected to no more than two other aromatic rings. 
An alternant PAH that is not a 1D alternant PAH is defined as a two-dimensional (2D) alternant PAH. 

To obtain the ground states of the alternant PAHs studied, TAO-LDA (i.e., TAO-DFT with the local density approximation) \cite{TAO-DFT} is employed for the lowest singlet and triplet energies of 
each alternant PAH on the respective geometries that were fully optimized at the same level of theory. The singlet-triplet energy gap (ST gap) of the alternant PAH is calculated as 
$(E_{\text{T}} - E_{\text{S}})$, the energy difference between the lowest triplet (T) and singlet (S) states. For the lowest singlet state of the alternant PAH, the active orbital occupation numbers 
obtained from TAO-LDA are adopted to assess the radical character of the alternant PAH. Here, the highest occupied molecular orbital (HOMO) is the ${(N/2)}^{\text{th}}$ orbital, and the lowest 
unoccupied molecular orbital (LUMO) is the ${(N/2 + 1)}^{\text{th}}$ orbital, with $N$ being the number of electrons in the alternant PAH. For all the TAO-LDA calculations, the optimal value of 
$\theta$ = 7 mhartree (as defined in Ref.\ \newcite{TAO-DFT}) is adopted. As the spin-restricted and spin-unrestricted TAO-LDA energies for the lowest singlet states of all the alternant PAHs 
adopted are essentially the same, and spin-unrestricted TAO-LDA calculations are always performed for the lowest triplet states, we remove the phrases ``spin-restricted" and ``spin-unrestricted" 
for brevity. 

All calculations are performed with a development version of \textsf{Q-Chem 4.0} \cite{Q-Chem}. Results are computed using the 6-31G(d) basis set with the fine grid EML(75,302), consisting of 
75 Euler-Maclaurin radial grid points and 302 Lebedev angular grid points.

\section*{Results and Discussion}

\subsection*{Validity of TAO-LDA Occupation Numbers} 

\begin{figure}[H] 
\includegraphics[scale=0.5]{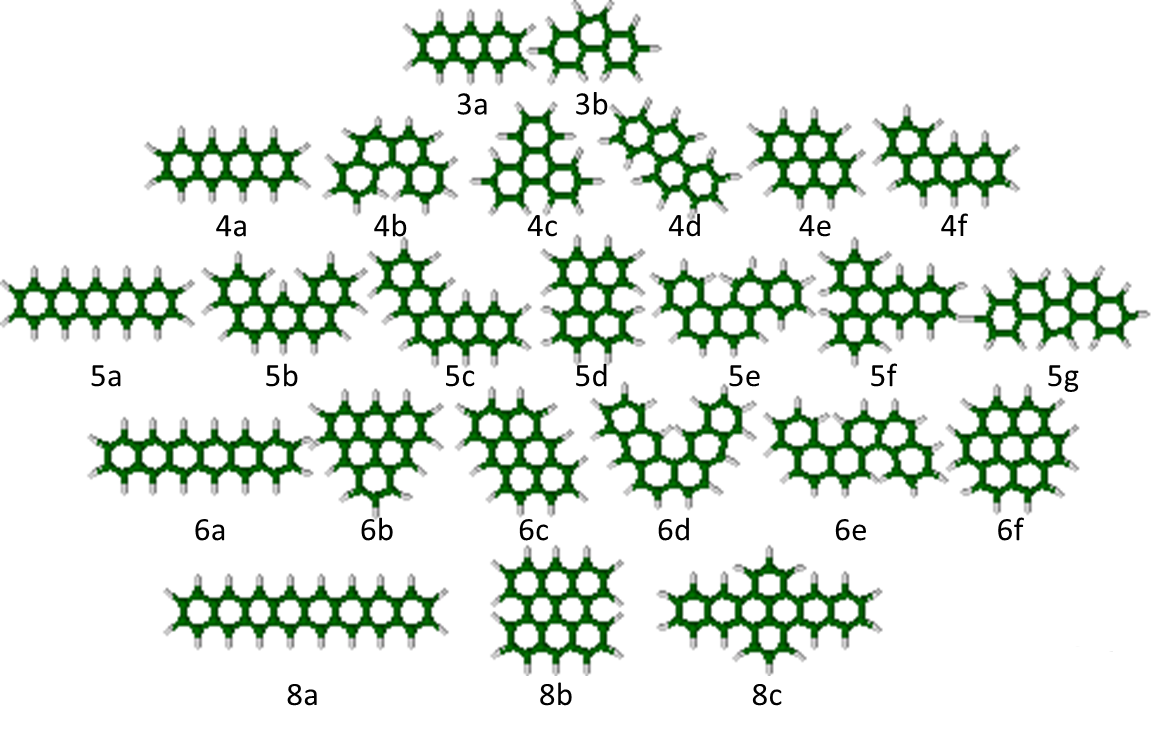} 
\caption{\label{test_molecules} 
Geometries of the 24 alternant PAHs studied.} 
\end{figure}

\begin{figure}[H] 
\includegraphics[scale=0.55]{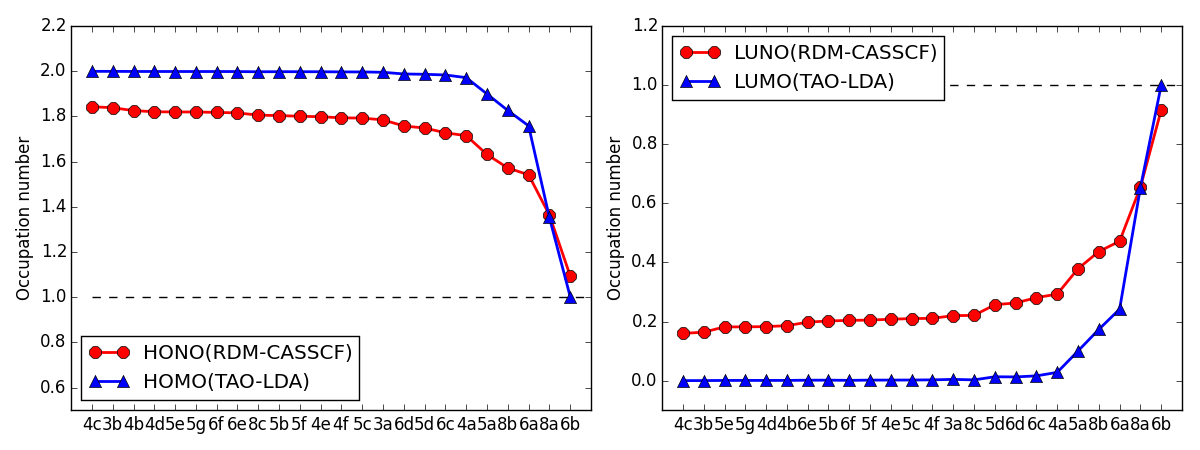} 
\caption{\label{comparison} 
Occupation numbers of the highest occupied ($n_\text{HONO}$) and lowest unoccupied ($n_\text{LUNO}$) natural orbitals obtained from the RDM-CASSCF method \cite{NOON1} and the 
occupation numbers of the highest occupied ($f_\text{HOMO}$) and lowest unoccupied ($f_\text{LUMO}$) molecular orbitals obtained from TAO-LDA, for the lowest singlet states of the 24 
alternant PAHs studied (see \Cref{test_molecules}).} 
\end{figure}

Here we examine the validity of TAO-LDA occupation numbers with respect to the NOONs obtained from the accurate RDM-CASSCF method \cite{NOON1}. As illustrated in \Cref{test_molecules}, we 
adopt a test set of 24 alternant PAHs, which are the PAHs (excluding the non-alternant PAHs) studied by Pelzer \emph{et al.} \cite{NOON1}. Based on the TAO-LDA calculations (see \Cref{table1}), 
the ST gaps of the 24 alternant PAHs are positive, so they all possess singlet ground states. Besides, the occupation numbers of the highest occupied ($f_\text{HOMO}$) and lowest unoccupied 
($f_\text{LUMO}$) molecular orbitals for the lowest singlet states (i.e., the ground states) of the 24 alternant PAHs, calculated using TAO-LDA are compared with the corresponding NOONs 
($n_\text{HONO}$ and $n_\text{LUNO}$, respectively) obtained from RDM-CASSCF \cite{NOON1}. Owing to the limitation of TAO-LDA (with a system-independent $\theta$ = 7 mhartree), the 
TAO-LDA occupation numbers are slightly biased towards closed-shell systems. A system-dependent $\theta$ (related to the distribution of NOONs) is expected to improve the general performance of 
TAO-LDA \cite{TAO-DFT}. Nevertheless, as shown in \Cref{comparison}, even with a fixed value of $\theta$ (i.e., 7 mhartree), the TAO-LDA occupation numbers are qualitatively similar to the 
RDM-CASSCF NOONs, yielding a similar trend for the radical character of the 24 alternant PAHs. Our TAO-LDA results support previous findings on the increase in radical character with increasing system 
size. For the alternant PAHs with the same number of aromatic rings, the geometrical arrangements of aromatic rings are shown to be responsible for the radical character, and it is observed that the smaller 
the ST gap, the stronger the radical character. Owing to its computational efficiency and reasonable accuracy, in this work, we adopt the TAO-LDA occupation numbers as the approximate NOONs to assess 
the radical character of the alternant PAHs studied.

\subsection*{1D Alternant PAHs} 

\begin{figure}[H] 
\includegraphics[scale=0.8]{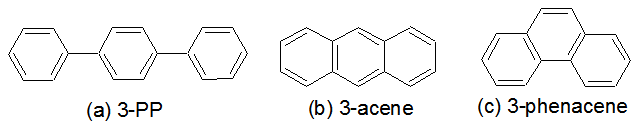} 
\caption{\label{1-D} 
Three possible ways to arrange the aromatic rings in 1D alternant PAHs.} 
\end{figure}

\Cref{1-D} represents three possible ways to arrange the aromatic rings in 1D alternant PAHs. Here the alternant PAH with armchair edges is denoted as $n$-PP [a planar poly(p-phenylene) oligomer] 
\cite{n-PP}, the alternant PAH with zigzag edges is denoted as $n$-acene, and the alternant PAH with the sawtooth arrangements of aromatic rings is denoted as $n$-phenacene, where $n$ is the number 
of aromatic rings in the alternant PAH. Based on the TAO-LDA calculations, $n$-PP, $n$-acene, and $n$-phenacene ($n$ = 3$-$20) all possess singlet ground states (see \Cref{table2}). 

For $n$-PP, the adjacent aromatic rings are connected to each other by a single C-C bond, which isolates each ring as if $n$-PP is simply the combination of several isolated benzenes. As each ring forms 
an aromatic sextet, $n$-PP does not have the non-Kekul\'{e} structures, and hence should display non-radical character. 

By contrast, for $n$-acene and $n$-phenacene, the adjacent aromatic rings are connected to each other by sharing one common side. Due to the sawtooth arrangements of aromatic rings, it is impossible 
to draw the non-Kekul\'{e} structures of $n$-phenacene. Therefore, $n$-phenacene should exhibit non-radical character, showing consistency with the findings of Plasser \emph{et al.} \cite{GNRs-MRAQCC}. 

\begin{figure}[H] 
\includegraphics[scale=0.7]{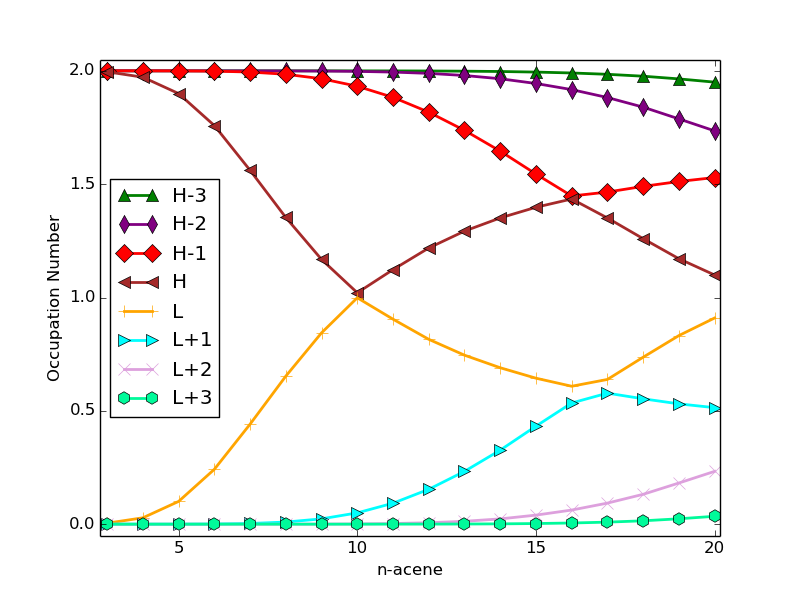} 
\caption{\label{acenes} 
Active orbital occupation numbers ($f_{\text{HOMO}-3}$, $f_{\text{HOMO}-2}$, $f_{\text{HOMO}-1}$, $f_{\text{HOMO}}$, $f_{\text{LUMO}}$, $f_{\text{LUMO}+1}$, $f_{\text{LUMO}+2}$, and 
$f_{\text{LUMO}+3}$) for the lowest singlet states of $n$-acenes as a function of the acene length $n$, calculated using TAO-LDA.} 
\end{figure}

On the other hand, regardless of the length of $n$-acene, the most important structure is the Kekul\'{e} structure with one aromatic sextet, the second most important structure is the non-Kekul\'{e} structure 
with two aromatic sextets and two unpaired $\pi$-electrons, and the third most important structure is the non-Kekul\'{e} structure with four aromatic sextets and four unpaired $\pi$-electrons. From the energy 
point of view (i.e., based on the previous models \cite{NK1,NK2,NK3}), the non-Kekul\'{e} structures can never be dominant, and hence $n$-acene should always display non-radical character (i.e., governed 
by the most important Kekul\'{e} structure). However, as shown in \Cref{table2a}, with the increase of $n$, the degeneracies of non-Kekul\'{e} structures (e.g., the second most important structure, the third 
most important structure, etc.) increase more rapidly than the degeneracy of the most important Kekul\'{e} structure. Therefore, based on our simple model, the weights of non-Kekul\'{e} structures should be 
increased with $n$, and may eventually be dominant for a large $n$. For a sufficiently large $n$, as the number of important non-Kekul\'{e} structures is increased, $n$-acene should display increasing 
polyradical character. As shown in \Cref{acenes}, the number of fractionally occupied orbitals obtained from TAO-LDA increases with the acene length, which are indeed in support of our argument. 
Therefore, in addition to the energies of the structures, the degeneracies of the structures should also be taken into account in the extended Clar's rule to qualitatively describe the polyradical character of 
$n$-acene \cite{Acene-DMRG,Acene-Jiang,GNRs-DMRG,GNRs-PHF,GNRs-MRAQCC,GNRs-MRAQCC2,TAO-DFT,TAO-GGA,GNRs-TAO}. 

Therefore, based on our simple model, for a given number of aromatic rings $n$, the strength of radical character in the three alternant PAHs is as follows: $n$-acene $>n$-phenacene $\approx$ $n$-PP. 
To verify this, the active orbital occupation numbers ($f_\text{HOMO}$ and $f_\text{LUMO}$) for the lowest singlet states of these alternant PAHs calculated using TAO-LDA are shown in \Cref{table2}. Our 
results suggest that $n$-acene should possess radical character for $n \ge 6$. By contrast, $n$-phenacene and $n$-PP exhibit non-radical character (even up to $n=20$). 

Based on our definition of 1D alternant PAHs, we can represent 1D alternant PAHs by a combination of the three types of arrangements in \Cref{1-D}, and qualitatively describe their radical character. Based 
on the arguments above, when the adjacent aromatic rings are connected to each other by a single C-C bond (e.g., $n$-PP), the alternant PAH exhibits non-radical character, regardless of the number of 
these rings. By contrast, when the adjacent aromatic rings are connected to each other by sharing one common side in the linear arrangement (e.g., $n$-acene), the degeneracies of the non-Kekul\'{e} 
structures are quickly increased with the number of these rings, displaying an increasing polyradical character. However, when the adjacent aromatic rings are connected to each other by sharing one 
common side in the sawtooth arrangement (e.g., $n$-phenacene), the number of aromatic sextets in the most important Kekul\'{e} structure is increased with the number of these rings, stabilizing the 
alternant PAH. Therefore, for a given number of aromatic rings $n$, the more the 1D alternant PAH resembles $n$-acene, the more it displays radical character. For example, the 1D alternant PAH with 
zigzag edges is much less stable than that with armchair edges \cite{GNR_new1,GNRs-TAO,n-PP,zig1,zig2,zig3}. 

\begin{figure}[H] 
\includegraphics[scale=0.8]{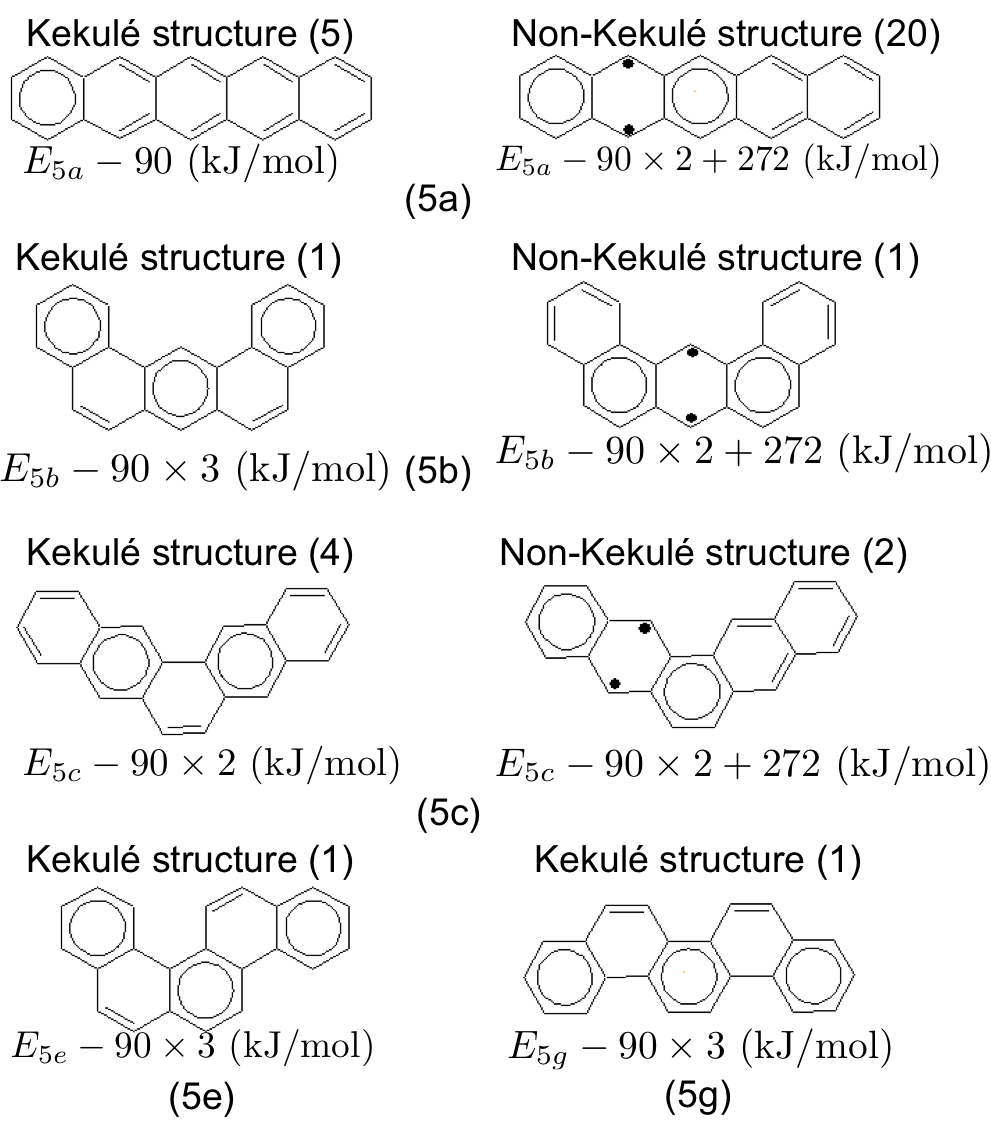} 
\caption{\label{1-D2} 
Most important Kekul\'{e} and non-Kekul\'{e} structures of 5a, 5b, 5c, 5e, and 5g. 
Note that 5e and 5g do not have the non-Kekul\'{e} structures. 
The energies and degeneracies (in parentheses) of the structures are shown.} 
\end{figure}

Here we take the five 1D alternant PAHs (5a, 5b, 5c, 5e, and 5g) in \Cref{test_molecules} as examples. Note that the other 5-ring alternant PAHs (5d and 5f) belong to 2D alternant PAHs. To assess their 
radical character, it is essential to estimate the weights of the most important non-Kekul\'{e} structures relative to those of the most important Kekul\'{e} structures (see \Cref{1-D2}). First, we recognize (5a) 
[the linear arrangement] and (5g) [the sawtooth arrangement] as the molecules with the most and least radical character, respectively. Secondly, the strength of radical character in (5e) should be the same 
as that in (5g), as (5e) and (5g) both have the most important Kekul\'{e} structures with three aromatic sextets, and do not have the non-Kekul\'{e} structures. Thirdly, for (5b) and (5c), while they both have 
the most important non-Kekul\'{e} structures with two aromatic sextets and two unpaired $\pi$-electrons, the Kekul\'{e} structure of (5b) is much more stable than that of (5c) [due to an extra aromatic sextet 
included in the Kekul\'{e} structure of (5b)]. Therefore, the weight of the non-Kekul\'{e} structure of (5c) should be greater than that of (5b), yielding stronger radical character for (5c). Finally, based on our 
simple model, the strength of radical character in the five 1D alternant PAHs is as follows: $(5a) > (5c) > (5b) > (5e) = (5g)$, matching reasonably well with the analysis of the TAO-LDA and RDM-CASSCF 
occupation numbers (see \Cref{table1}). Note that these alternant PAHs are far away from being governed by the non-Kekul\'{e} structures, and the difference of radical character among them is small. 
However, we can still qualitatively describe and compare the strength of radical character in these alternant PAHs by simply studying the most important Kekul\'{e} and non-Kekul\'{e} structures in our simple 
model.

\subsection*{2D Alternant PAHs} 

For 2D alternant PAHs, Pelzer \emph{et al.} argued that the dimensionality may be an effective predictor when comparing 1D PAHs to 2D PAHs, in the sense that 1D PAHs generally exhibit more radical 
character than 2D PAHs \cite{NOON1}. However, as they commented, the above argument alone cannot explain why the longer, narrower structure (8c) exhibits less radical character than the square-shaped 
structure (8b) [see \Cref{table1}]. By contrast, based on our simple model, the geometry of (8c) allows more aromatic sextets to be included in the most important Kekul\'{e} structure, so the radical character 
of (8c) is less governed by the most important non-Kekul\'{e} structure. Therefore, (8c) exhibits less radical character than (8b). This example highlights the importance of studying the Kekul\'{e} and 
non-Kekul\'{e} structures in the extended Clar's rule for qualitatively describing the radical character of alternant PAHs. 

\begin{figure}[H] 
\includegraphics[scale=0.8]{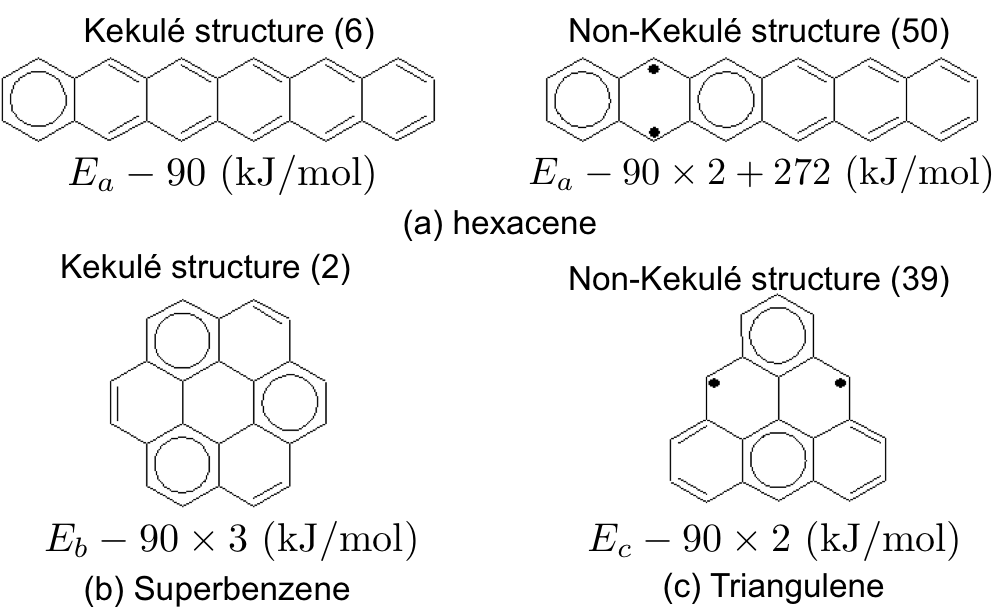} 
\caption{\label{2-D} 
Most important Kekul\'{e} and non-Kekul\'{e} structures of hexacene, superbenzene, and triangulene. 
Note that superbenzene does not have the non-Kekul\'{e} structures, and triangulene does not have the Kekul\'{e} structures. 
The energies and degeneracies (in parentheses) of the structures are shown.} 
\end{figure}

Similar arguments may be applied to the three alternant PAHs in \Cref{example}. \Cref{2-D} shows the most important Kekul\'{e} and non-Kekul\'{e} structures of hexacene, superbenzene, and triangulene. 
Systems with greater symmetry (e.g., superbenzene) do not necessarily exhibit stronger radical character \cite{NOON1}. As superbenzene does not have the non-Kekul\'{e} structures, and hence the 
Kekul\'{e} structures must be dominant, superbeneze exhibits non-radical character. By contrast, it is impossible to draw the Kekul\'{e} structures of triangulene, and two unpaired $\pi$-electrons always 
appear in the most important non-Kekul\'{e} structures, yielding the diradical character. 

\begin{figure}[H] 
\includegraphics[scale=0.8]{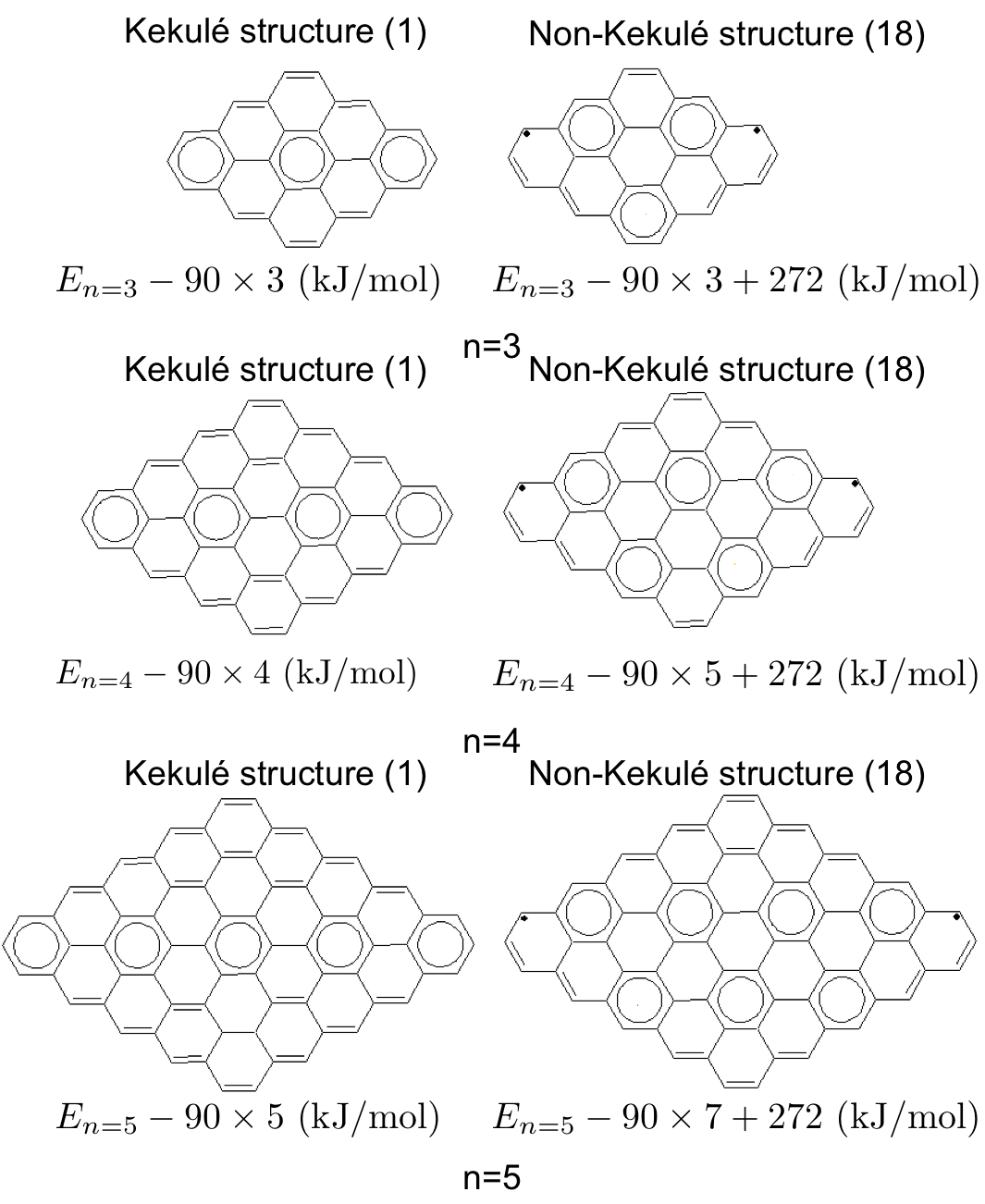} 
\caption{\label{2-D3} 
Most important Kekul\'{e} and non-Kekul\'{e} structures of zigzag-edged diamond-shaped graphene nanoflakes (C$_{2n^{2}+6n+1}$H$_{4n+2}$) with $n$ = 3, 4, and 5. 
The energies and degeneracies (in parentheses) of the structures are shown.} 
\end{figure}

In contrast to 1D alternant PAHs, it is not practical to classify 2D alternant PAHs into groups, and formulate general rules describing their radical character, as there are too many possible geometrical 
arrangements of aromatic rings to study all of them. In this work, we adopt the following two types of 2D alternant PAHs: zigzag-edged triangular graphene nanoflakes (C$_{n^{2}+4n+1}$H$_{3n+3}$) 
[see \Cref{2-D2}] and zigzag-edged diamond-shaped graphene nanoflakes (C$_{2n^{2}+6n+1}$H$_{4n+2}$) [see \Cref{2-D3}], with $n$ being the number of aromatic rings at each side, as the test systems 
to investigate the role of Kekul\'{e} and non-Kekul\'{e} structures in the radical character of these 2D alternant PAHs. 

\subsubsection*{Zigzag-Edged Triangular Graphene Nanoflakes} 

\begin{figure}[H] 
\includegraphics[scale=0.8]{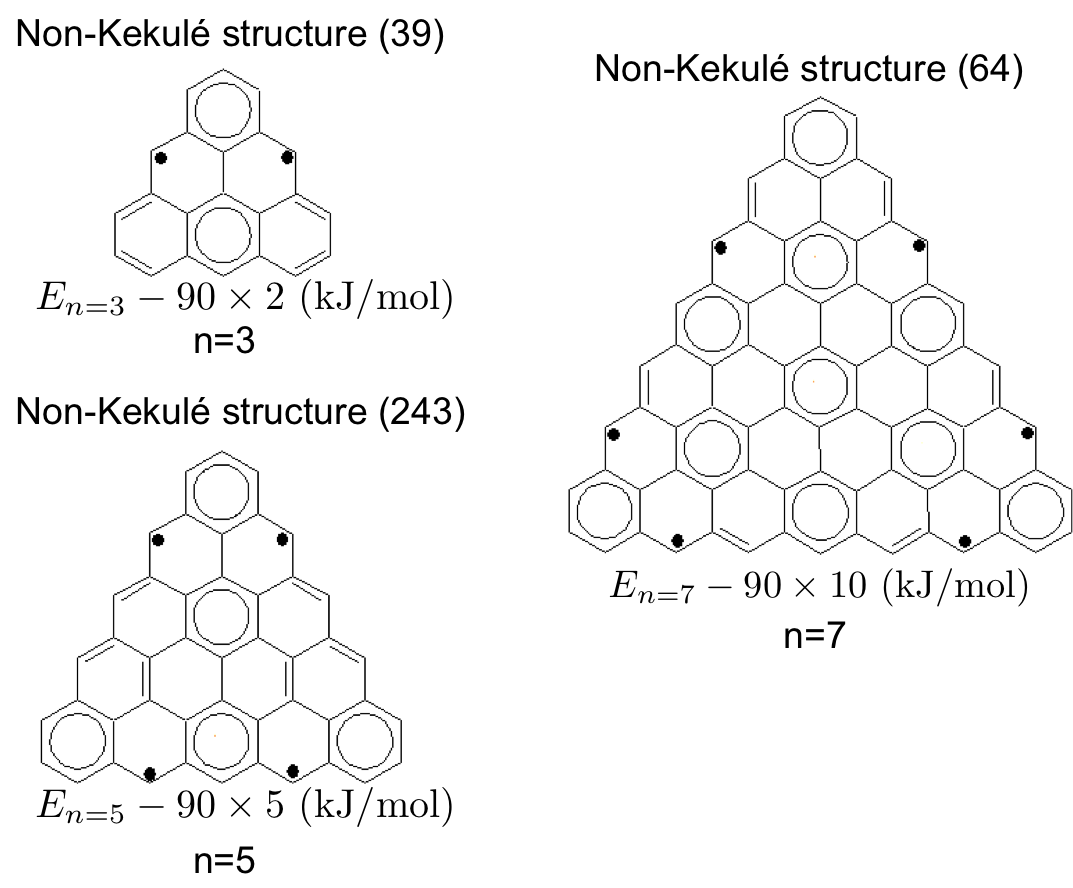} 
\caption{\label{2-D2} 
Most important non-Kekul\'{e} structures of zigzag-edged triangular graphene nanoflakes (C$_{n^{2}+4n+1}$H$_{3n+3}$) with $n$ = 3, 5, and 7. 
Note that zigzag-edged triangular graphene nanoflakes do not have the Kekul\'{e} structures. 
The energies and degeneracies (in parentheses) of the structures are shown.} 
\end{figure}

Zigzag-edged triangular graphene nanoflakes (C$_{n^{2}+4n+1}$H$_{3n+3}$) with $n$ = 3, 5, 7, 9, and 11, all possess singlet ground states, based on the TAO-LDA calculations (see \Cref{table3}). Note 
that those with an even number of $n$ are excluded, as they contain an odd number of electrons, and do not possess singlet states. Similar to triangulene (i.e., $n=3$), zigzag-edged triangular graphene 
nanoflakes do not have the Kekul\'{e} structures, and hence the non-Kekul\'{e} structures must be dominant. \Cref{2-D2} shows the most important non-Kekul\'{e} structures of zigzag-edged triangular 
graphene nanoflakes with $n$ = 3, 5, and 7. Due to the peculiar geometries of zigzag-edged triangular graphene nanoflakes, unpaired $\pi$-electrons always appear in the most important non-Kekul\'{e} 
structures (as well as the standard structures). According to \Cref{2-D2}, there are 2, 4, and 6 unpaired $\pi$-electrons for the $n$ = 3, 5, and 7 cases, respectively. We can extend this observation, and 
argue that there should be ($n-1$) unpaired $\pi$-electrons in the zigzag-edged triangular graphene nanoflake with $n$ aromatic rings at each side. As shown in \Cref{table3}, the active orbital occupation 
numbers for the lowest singlet states of zigzag-edged triangular graphene nanoflakes with $n$ = 3, 5, 7, 9, and 11, calculated using TAO-LDA, are indeed in support of our simple model. Therefore, 
zigzag-edged triangular graphene nanoflakes with longer side length should exhibit increasing polyradical character. 

\subsubsection*{Zigzag-Edged Diamond-Shaped Graphene Nanoflakes} 

Zigzag-edged diamond-shaped graphene nanoflakes (C$_{2n^{2}+6n+1}$H$_{4n+2}$) with $n$ = 3$-$7, all have singlet ground states, based on the TAO-LDA calculations (see \Cref{table4}). \Cref{2-D3} 
shows the most important Kekul\'{e} and non-Kekul\'{e} structures of zigzag-edged diamond-shaped graphene nanoflakes with $n$ = 3$-$5 and their corresponding degeneracies. For $n=3$, as the energy 
of the most important Kekul\'{e} structure is lower than that of the most important non-Kekul\'{e} structure, and the degeneracy of the most important non-Kekul\'{e} structure is insufficiently large, the radical 
character of the molecule is mainly governed by the most important Kekul\'{e} structure, hence displaying a closed-shell singlet ground state. For the larger $n$, while the degeneracies of the most important 
Kekul\'{e} and non-Kekul\'{e} structures remain the same as those for $n=3$, the energy difference between these two structures becomes smaller, and hence the weight of non-Kekul\'{e} structure becomes 
relatively larger. For $n > 6$, the energy of the most important non-Kekul\'{e} structure becomes even lower than that of the most important Kekul\'{e} structure, so the molecule displays strong radical 
character. As shown in \Cref{table4}, the active orbital occupation numbers for the lowest singlet states of zigzag-edged diamond-shaped graphene nanoflakes with $n$ = 3$-$7, calculated using TAO-LDA, 
show consistency with our simple model.

\section*{Conclusions} 

In conclusion, we have shown that the TAO-LDA occupation numbers are qualitatively similar to the NOONs obtained from the accurate RDM-CASSCF method, and are potentially useful for assessing the 
radical character of large alternant PAHs, due to its computational efficiency. Relative to the analysis of the TAO-LDA and RDM-CASSCF occupation numbers, the studies of Kekul\'{e} and non-Kekul\'{e} 
structures in our proposed simple model qualitatively describe the radical character of alternant PAHs, which could be useful when electronic structure calculations are infeasible due to the expensive 
computational cost. Our results support previous findings on the increase in radical character with increasing system size. For alternant PAHs with the same number of aromatic rings, the geometrical 
arrangements of aromatic rings have been shown to be influential for the radical character of alternant PAHs. For 1D alternant PAHs, the more the molecules resemble the acene series, the more they 
display polyradical character. For 2D alternant PAHs, it is impractical to classify the molecules into groups, and formulate general rules describing their radical character, as there are too many possible 
geometrical arrangements of aromatic rings to study all of them. Nevertheless, the radical character of several 2D alternant PAHs has been qualitatively described by our simple model.

\begin{acknowledgement} 

This work was supported by the Ministry of Science and Technology of Taiwan (Grant No.\ MOST104-2628-M-002-011-MY3), National Taiwan University (Grant No.\ NTU-CDP-105R7818), 
the Center for Quantum Science and Engineering at NTU (Subproject Nos.:\ NTU-ERP-105R891401 and NTU-ERP-105R891403), and the National Center for Theoretical Sciences of Taiwan. 

\end{acknowledgement}

\section*{Author Contributions} 
C.-N.Y. and J.-D.C. designed the project. C.-N.Y. performed the calculations. C.-N.Y. and J.-D.C. contributed to the data analysis, and wrote the paper.

\section*{Additional Information} 
{\bf Competing financial interests:} The authors declare no competing financial interests.

\newpage 

\newpage 

\newpage 

\newpage 

\newpage 

\newpage 

\newpage 

\newpage 

\newpage 

\newpage 

\newpage 

\newpage 

\newpage 
\begin{table} 
\caption{\label{table1} 
Active orbital occupation numbers ($f_\text{HOMO}$ and $f_\text{LUMO}$) for the lowest singlet states of the 24 alternant PAHs studied (see \Cref{test_molecules}), calculated using TAO-LDA. 
For comparison, the corresponding natural orbital occupation numbers ($n_\text{HONO}$ and $n_\text{LUNO}$, respectively) obtained from the RDM-CASSCF method \cite{NOON1} are shown. 
As the singlet-triplet energy gaps $E_{\text{ST}}$ (in eV) of these alternant PAHs are positive, they all possess singlet ground states.} 
\begin{tabular}{c|cc|ccc} 
& RDM-CASSCF &&& TAO-LDA & \\
\hline 
& $n_\text{HONO}$ & $n_\text{LUNO}$ & $f_\text{HOMO}$ & $f_\text{LUMO}$ & $E_{\text{ST}}$ \\ 
\hline
3a&1.785&0.220&1.996&0.004&1.87\\
3b&1.839&0.164&2.000&0.000&2.74\\
\hline
4a&1.715&0.293&1.972&0.028&1.26\\
4f&1.794&0.211&1.997&0.003&2.04\\
4e&1.799&0.208&1.998&0.002&2.17\\
4d&1.821&0.183&1.999&0.001&2.45\\
4b&1.826&0.186&1.999&0.001&2.46\\
4c&1.843&0.160&2.000&0.000&2.82\\
\hline
5a&1.632&0.379&1.900&0.100&0.85\\
5d&1.749&0.257&1.987&0.013&1.54\\
5c&1.793&0.210&1.997&0.002&1.86\\
5f&1.801&0.205&1.998&0.002&2.15\\
5b&1.804&0.202&1.998&0.002&2.16\\
5e&1.820&0.182&1.999&0.001&2.34\\
5g&1.820&0.182&1.999&0.001&2.35\\
\hline 
6b&1.095&0.916&1.000&1.000&0.28\\
6a&1.542&0.472&1.759&0.242&0.59\\
6c&1.728&0.281&1.984&0.016&1.47\\
6d&1.758&0.263&1.988&0.013&1.56\\
6e&1.816&0.198&1.999&0.002&2.23\\
6f&1.818&0.204&1.999&0.001&2.23\\
\hline
8a&1.364&0.655&1.352&0.653&0.34\\
8b&1.572&0.437&1.827&0.173&0.72\\
8c&1.806&0.221&1.998&0.003&1.87\\
\end{tabular}
\end{table} 

\newpage 
\begin{table} 
\caption{\label{table2} 
Active orbital occupation numbers ($f_\text{HOMO}$ and $f_\text{LUMO}$) for the lowest singlet states of $n$-PP, $n$-acene, and $n$-phenacene ($n$ = 3$-$20), calculated using TAO-LDA. 
As the singlet-triplet energy gaps $E_{\text{ST}}$ (in eV) of these alternant PAHs are positive, they all possess singlet ground states.} 
\begin{tabular}{c|ccc|ccc|ccc}
&& $n$-PP &&& $n$-acene &&& $n$-phenacene & \\ 
\hline 
$n$ & $f_\text{HOMO}$ & $f_\text{LUMO}$ & $E_{\text{ST}}$ & $f_\text{HOMO}$ & $f_\text{LUMO}$ & $E_{\text{ST}}$ & $f_\text{HOMO}$ & $f_\text{LUMO}$ & $E_{\text{ST}}$ \\ 
\hline
3&1.999&0.001&2.51&1.996&0.004&1.87&2.000&0.002&2.74\\
4&1.998&0.002&2.24&1.972&0.028&1.26&1.999&0.001&2.45\\
5&1.997&0.003&2.06&1.900&0.100&0.85&1.999&0.001&2.35\\
6&1.996&0.004&1.93&1.759&0.242&0.59&1.999&0.001&2.25\\
7&1.995&0.006&1.83&1.561&0.441&0.43&1.998&0.001&2.17\\
8&1.994&0.007&1.74&1.353&0.653&0.34&1.998&0.001&2.11\\
9&1.993&0.008&1.67&1.168&0.844&0.29&1.998&0.001&2.05\\
10&1.992&0.008&1.62&1.020&1.000&0.26&1.998&0.002&2.00\\
11&1.992&0.009&1.57&1.123&0.904&0.23&1.998&0.002&1.96\\
12&1.991&0.010&1.52&1.218&0.816&0.21&1.998&0.002&1.92\\
13&1.991&0.010&1.48&1.293&0.746&0.19&1.998&0.002&1.89\\
14&1.990&0.010&1.45&1.351&0.690&0.17&1.997&0.002&1.85\\
15&1.990&0.011&1.42&1.397&0.645&0.16&1.997&0.002&1.83\\
16&1.990&0.011&1.39&1.435&0.608&0.15&1.997&0.002&1.80\\
17&1.989&0.011&1.36&1.351&0.638&0.14&1.997&0.002&1.77\\
18&1.989&0.012&1.33&1.259&0.737&0.13&1.997&0.002&1.75\\
19&1.989&0.012&1.31&1.176&0.827&0.12&1.997&0.002&1.73\\
20&1.989&0.012&1.29&1.099&0.910&0.11&1.997&0.002&1.71\\
\end{tabular}
\end{table}

\newpage 
\begin{table} 
\caption{\label{table2a} 
Degeneracies of the top three most important structures of $n$-acene as a function of the acene length $n$. Here 
the most important structure is the Kekul\'{e} structure with one aromatic sextet, 
the second most important structure is the non-Kekul\'{e} structure with two aromatic sextets and two unpaired $\pi$-electrons, and 
the third most important structure is the non-Kekul\'{e} structure with four aromatic sextets and four unpaired $\pi$-electrons.} 
\begin{tabular}{c|c|c|c} 
$n$ & 
Most important & Second most important & Third most important \\ 
\hline
3&3&1&0\\
4&4&6&0\\
5&5&20&1\\
6&6&50&10\\
7&7&105&53\\
8&8&196&200\\
9&9&336&606\\
10&10&540&1572\\
11&11&825&3630\\
12&12&1210&7656\\
13&13&1716&15015\\
14&14&2366&27742\\
15&15&3185&48763\\
16&16&4200&82160\\
17&17&5440&133484\\
18&18&6936&209924\\
19&19&8721&321091\\
20&20&10830&479586\\
\end{tabular}
\end{table}

\newpage 
\begin{table} 
\caption{\label{table3} 
Active orbital occupation numbers ($f_{\text{HOMO}-5}$, ..., $f_{\text{HOMO}-1}$, $f_{\text{HOMO}}$, $f_{\text{LUMO}}$, $f_{\text{LUMO}+1}$, ..., and $f_{\text{LUMO}+5}$) 
for the lowest singlet states of zigzag-edged triangular graphene nanoflakes (C$_{n^{2}+4n+1}$H$_{3n+3}$) as a function of the side length $n$, calculated using TAO-LDA. 
As the singlet-triplet energy gaps $E_{\text{ST}}$ (in eV) of these alternant PAHs are positive, they all possess singlet ground states.} 
\begin{tabular}{c|c|c|c|c|c}
$n$ &3&5&7&9&11\\
\hline
$f_{\text{HOMO}-5}$&2.000&2.000&2.000&1.998&1.995\\ 
$f_{\text{HOMO}-4}$&2.000&2.000&2.000&1.998&1.105\\ 
$f_{\text{HOMO}-3}$&2.000&2.000&2.000&1.097&1.105\\ 
$f_{\text{HOMO}-2}$&2.000&2.000&1.082&1.097&1.092\\ 
$f_{\text{HOMO}-1}$&2.000&1.055&1.082&1.076&1.029\\ 
$f_{\text{HOMO}}$&1.000&1.055&1.045&0.998&0.999\\ 
$f_{\text{LUMO}}$&1.000&0.981&0.963&0.962&0.999\\ 
$f_{\text{LUMO}+1}$&0.000&0.981&0.963&0.962&0.929\\ 
$f_{\text{LUMO}+2}$&0.000&0.000&0.914&0.904&0.929\\ 
$f_{\text{LUMO}+3}$&0.000&0.000&0.000&0.904&0.914\\ 
$f_{\text{LUMO}+4}$&0.000&0.000&0.000&0.001&0.901\\ 
$f_{\text{LUMO}+5}$&0.000&0.000&0.000&0.001&0.002\\ 
\hline
$E_{\text{ST}}$ &0.28&0.08&0.05&0.04&0.03 \\ 
\end{tabular} 
\end{table} 

\newpage 
\begin{table} 
\caption{\label{table4} 
Active orbital occupation numbers ($f_{\text{HOMO}-5}$, ..., $f_{\text{HOMO}-1}$, $f_{\text{HOMO}}$, $f_{\text{LUMO}}$, $f_{\text{LUMO}+1}$, ..., and $f_{\text{LUMO}+5}$) 
for the lowest singlet states of zigzag-edged diamond-shaped graphene nanoflakes (C$_{2n^{2}+6n+1}$H$_{4n+2}$) as a function of the side length $n$, calculated using TAO-LDA. 
As the singlet-triplet energy gaps $E_{\text{ST}}$ (in eV) of these alternant PAHs are positive, they all possess singlet ground states.} 
\begin{tabular}{c|c|c|c|c|c} 
$n$ &3&4&5&6&7\\
\hline 
$f_{\text{HOMO}-5}$&2.000&2.000&2.000&2.000&1.999\\ 
$f_{\text{HOMO}-4}$&2.000&2.000&2.000&2.000&1.997\\ 
$f_{\text{HOMO}-3}$&2.000&2.000&2.000&1.995&1.950\\ 
$f_{\text{HOMO}-2}$&2.000&2.000&1.990&1.905&1.672\\ 
$f_{\text{HOMO}-1}$&1.999&1.976&1.821&1.523&1.277\\ 
$f_{\text{HOMO}}$&1.938&1.668&1.351&1.161&1.070\\
$f_{\text{LUMO}}$&0.062&0.334&0.660&0.853&0.946\\
$f_{\text{LUMO}+1}$&0.000&0.021&0.170&0.479&0.745\\
$f_{\text{LUMO}+2}$&0.000&0.000&0.008&0.080&0.300\\
$f_{\text{LUMO}+3}$&0.000&0.000&0.000&0.004&0.040\\
$f_{\text{LUMO}+4}$&0.000&0.000&0.000&0.000&0.002\\
$f_{\text{LUMO}+5}$&0.000&0.000&0.000&0.000&0.001\\
\hline
$E_{\text{ST}}$ &1.02&0.44&0.20&0.12&0.08 \\ 
\end{tabular} 
\end{table} 

\end{document}